# Black holes and the supermassive compact object at the Galactic center: multi-arts of thought and nature

Qingjuan Yu

Kavli Institute for Astronomy and Astrophysics, and School of Physics, Peking University, 100871 Beijing, China
Email: yuqj@pku.edu.cn

The Nobel Prize in Physics 2020 was awarded to Roger Penrose "for the discovery that black hole formation is a robust prediction of the general theory of relativity," and Reinhard Genzel and Andrea Ghez "for the discovery of a supermassive compact object at the center of our galaxy." A review of the awarded works presents us with an inspiring exploration history of more than 200 years toward understanding one of the most enigmatic objects in the spacetime, black holes (BHs), providing challenges to the most intelligent minds and the most advanced experiments of our time. These works are the multi-arts of astronomy, physics, mathematics, technology, and nature.

Existence and universality, the two basic characteristics of questioning everything in the universe, have been incorporated into understanding not only BHs in mathematical physics but also supermassive compact objects (COs) at the centers of galaxies (Figure 1).

John Michell (1783) and Pierre-Simon Laplace (1796–1799) introduced the object of "dark stars" in Newtonian spacetime, which have ordinary matter densities but are so massive that even light cannot escape from their gravitational potential. Albert Einstein's theory of general relativity (1915) revolutionized our view of spacetime, and its vacuum solutions of the Schwarzschild metric (1916) and the Kerr metric (1963) show the mathematic existence of singularity (at which no one knows how physical laws work) and event horizon (within which no communication with an outside observer is likely). In the framework of general relativity, Robert Oppenheimer and Hartland Snyder (1939) identified the formation of a horizon in studying purely spherical collapse of uniform dust cloud spheres and further the formation of a singularity. The above existence of a singularity and a horizon in mathematical physics and in the spherical collapse model are for highly symmetric systems (either spherical or axisymmetric), and it was not clear whether the singularity can form in reality with deviations from symmetry. Without presuming those symmetries, and by applying topology, Penrose's singularity theorem (1965)[1] dispelled doubts and showed the conditional universality of the collapse toward a singularity in a general system with a positive energy density as long as a "trapped surface" forms, where the existence of a trapped surface is independent of an assumption about symmetry. Note that the theorem does not mean that the formation of a trapped surface is a necessary condition for the collapse toward a singularity. The possibility for the existence of a "naked" singularity, which is also one part of the Kerr metric solution, is not excluded, although Penrose's cosmic censorship conjecture is used to forbid its existence.

Nature has provided us with great hints for the existence and universality of supermassive BHs in the real universe. Hints as to their existence can be dated back to the discovery of quasars in the 1960s, and the leading model proposed to explain the engine for the huge luminosities of quasars is gas accretion onto supermassive BHs in those distant galaxies.

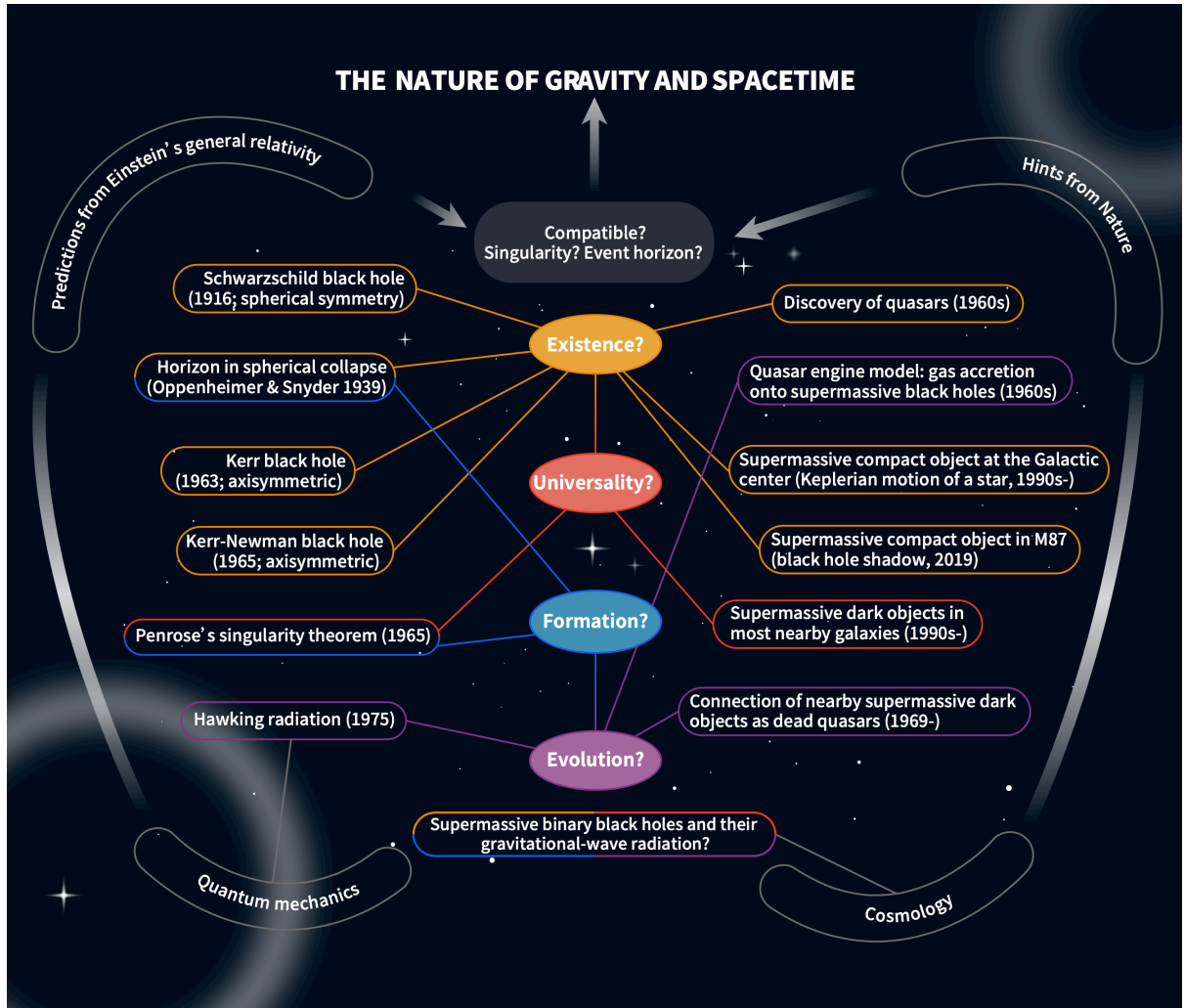

Figure 1: A Picture toward Understanding BHs and the Nature of Gravity and Spacetime. The diagram shows the backgrounds, developments, and relevance in the discovery of the singularity theorem and the discovery of the supermassive CO at the center of our galaxy (shortened as "Galactic center"). The connections between those discoveries need to be fulfilled on the road toward understanding BHs and the nature of gravity and spacetime.

According to the cosmological principle that no place in the universe is special in the large scale, distant galaxies can represent the history of the galaxies in our nearby universe, and it is expected that supermassive BHs exist as "dead quasars" in nearby quiescent galaxies.[2] Searches in the centers of nearby galaxies have confirmed the existence and the generality of supermassive COs. The direct connection of those objects existing in reality with the BHs in the mathematical solutions of general relativity and in the singularity theorem is breached by the unlikely direct detection of singularities and event horizons. However, it can be demonstrated that objects in reality are close to the expectations of those expected from a BH or any other possible objects.

One general constraint is the measurement of the mass density limit in the inner region of galactic nuclei. The center of our galaxy is unique due to its closeness to us, at a distance of 25,000 light years. The Keplerian motion of a star S2 (or S0-2) around the central compact radio source Sgr A∗ at the Galactic center, monitored by both the Genzel team and the Ghez team for more than two decades, provides a best constraint on the compactness of a central supermassive object. (The orbit is easily reminiscent of the Keplerian motions of the planets around the sun in our solar system.) The orbital period of S2 is about 16 years and its pericenter distance is only 125 AU from Sgr A∗. The mass of the central CO constrained from the Keplerian orbit is $4 \times 10^6$ $M_\odot$, and the constrained minimum mass density within the S2 orbit is $5 \times 10^{15}$ $M_\odot/pc^3$. This cannot be explained by any mass extension of any known astrophysical sources outside the S2 orbit, and an inner hypothesized dense dark star cluster with the same mass density would have a very short (evaporation) lifetime of ~$10^6$ years, and before that collapse into a BH is one possible destiny of the gravitational evolution of such a cluster. The mass centroid of the S2 orbit lies within 16 AU of Sgr A∗, which has an apparent size of less than 1 AU and no detectable proper motion. All the observations are compatible with the explanation of a supermassive CO from a supermassive BH, although not all exotic objects can be ruled out. The developments of the techniques of speckle imaging and adaptive optics in large telescopes play a key role in determining the motion of S2, and the compactness has been constrained by the prototype of the orbit in 2002-2003.[3,4] Some constraints from early observations in nearby galactic centers (e.g., $4 \times 10^9$ $M_\odot/pc^3$, i.e., $3.6 \times 10^7$ $M_\odot$ within a radius of 0.13 pc, through the Keplerian rotation of a gaseous disk in NGC 4258 in 1995[5]) provided a less compact constraint, and the lifetime of a hypothesized dense dark star cluster with the same mass density can be about $10^{10}$ years. The BH shadow image of M87 obtained in 2019[6] adds another case of a constraint on the high compactness of the supermassive object, $4 \times 10^{17}$ $M_\odot/pc^3$.

By analyzing the stellar velocity dispersion distributions at galactic centers, not limited to being in several specific individual galaxies, the supermassive COs are shown to ubiquitously exist in the spheroidal components of nearby galaxies,[7] although the connections to supermassive BHs are not as strong as the above individual cases because of the limited spatial resolution of the data used and hence the less constrained compactness of the objects. The demographics of supermassive COs has been further developed. However, it is not clear whether these supermassive dark objects in nearby galactic centers are really the remnants of dead quasars.

The connection between the supermassive dark objects in the nearby universe and the engines in distant quasars has been pinned down demographically.[8] With the mass-to-energy conversion efficiency predicted by gas accretion onto (rotating) Kerr BHs, the mass densities in the supermassive dark objects in the nearby universe are consistent with the mass density of gas accreted in quasar phases, which concludes that those dark objects are indeed supermassive BHs in dead quasars and that their mass originates mainly from gas accretion during the quasar phases.

Beyond existence and universality, formation and evolution are among the fundamental (dynamical) questions with regard to every species/object in the world (Figure 1): more information is needed to answer these questions. Penrose's singularity theorem touches on one aspect of the formation of singularities, but the correspondence in the real universe of the formation process and the necessity of the formation condition remain to be tested. The only specific case of a supermassive CO at the Galactic center cannot describe the entire history of the evolution of the object, as one individual source can only represent a specific time in the evolution. A census of these objects in the universe can cover different time periods of individual cases, and a statistical method is needed to study the evolution from "hints" from Nature. In addition, the roles of the environments or host galaxies of supermassive COs cannot be ignored.

The universality of supermassive BHs at galactic centers and the processes of galaxy mergers in hierarchical cosmology will naturally lead to questioning the existence of a BH companion at the Galactic center and the universality of supermassive binary BHs in other galaxies,[9] and further stimulate the exploration of their formation and evolution. The current observations do not rule out the existence of a BH companion at the Galactic center, and there are also numerous observational hints about binary BH candidates in other galaxies. The gravitational wave radiation from merging binary BHs at their last evolutionary stage provides a probe of the nature of gravity and spacetime.

The road toward filling the gap between the singularity theorem and the supermassive CO at the Galactic center is toward understanding the nature of gravity and spacetime ([Figure 1](#)). Does the existence in mathematics really exist in Nature? Does a singularity really exist, and can gravity and spacetime be quantized? What is the spacetime structure around the supermassive COs? Current and future instrument development is making the testing of general relativity and gravity theory in the strong field possible. Development of the GRAVITY instrument (with an interferometer) is making progress in detecting closer stars and infrared flares at three to five times of the Schwarzschild radius around Sgr A∗. The gravitational redshift and the Schwarzschild precession,[10] revealed by the S2 orbit, are compatible with the prediction of general relativity. Ongoing and planned ground and space instruments for detecting gravitational waves and BH shadows will potentially reveal spacetime close to the event horizon.

The 2020 Nobel Prize in Physics, and those in the recent years (2017 and 2019), remind us that the expansion of human view horizons beyond physics and the environment on Earth is becoming to be appreciated. We are in a great era to experience the transition of our view horizons. In the past 100 years, the creation of quantum mechanics and the expansion of humanity's horizon to the micro world has revolutionized our world view, and many masterpieces generated in the world of quantum physics were honored with Nobel prizes. Without doubt, more masterpieces toward understanding gravity and spacetime will be revealed with the expansion of our horizon to the extreme environments beyond Earth, and many more in the case of a successful fusion with the quantum world.

The works awarded the Nobel Prize in Physics in 2020 are rich in methodology, working styles, and viewing angles with regard to understanding these mysterious objects in the universe. Observational astronomical discoveries show that the theory of mathematical physics is alive and stimulate its development. The prediction of mathematical physics motivates observations and provides the goal or soul for observational discoveries. Individual work and team collaborations/competition all push our understandings forward. The exploration of the nature of spacetime is eternal. Along this road, it is fortunate to have environments that can nurture and encourage the curiosity for fundamental questions and the inclusive and diversity of innovation. It is fortunate to have the occasion to appreciate the art of logic and nature, and appreciate those steps toward understanding the universe, labeled or unlabeled with Nobel Prizes, every year.

Acknowledgments

This work was supported in part by the National Natural Science Foundation of China under nos. 11673001 and 11721303, and the National Key R&D Program of China (grant no. 2016YFA0400703).